\documentclass[aps, prl, groupedaddress, amsmath,
amssymb, floatfix, reprint, twocolumn,showpacs]{revtex4-1}

\usepackage{graphicx}
\usepackage{textcomp}
\usepackage{tikz}
\usepackage{bm}
\usepackage{dcolumn}
\usepackage{booktabs}
\usepackage[utf8]{inputenc}
\usepackage[colorlinks=true]{hyperref}
\usepackage[scaled]{helvet}
\usepackage{sansmath}

\graphicspath{{./figures/}}

\newcommand{\ket}[1]{\vert#1\rangle}

\newcommand{\NdYSO}[0]{Nd$^{3+}$:Y$_2$SiO$_5$}

\newcommand{\bluecircle}[0]{%
  \begin{tikzpicture}[x=1pt, y=1pt]
    \draw [fill=blue, thick] (0, -2) circle (2.5);
  \end{tikzpicture}%
}

\definecolor{darkred}{rgb}{0.545, 0.0, 0.0}
\newcommand{\redsquare}[0]{%
  \begin{tikzpicture}[x=1pt, y=1pt]
    \draw [darkred, very thick] rectangle (4.5, 4.5);
  \end{tikzpicture}%
}

\begin{document}

\title{Quantum storage of polarization qubits in birefringent and
  anisotropically absorbing materials}

\author{Christoph~Clausen}
\author{F\'{e}lix Bussi\`{e}res}
\author{Mikael~Afzelius}
\email{mikael.afzelius@unige.ch} 
\author{Nicolas Gisin}
\affiliation{Group of Applied Physics, University of Geneva, CH-1211
  Geneva 4, Switzerland}

\date{\today}

\begin{abstract}
  Storage of quantum information encoded into true single photons is
  an essential constituent of long-distance quantum communication
  based on quantum repeaters and of optical quantum information
  processing. The storage of photonic polarization qubits is, however,
  complicated by the fact that many materials are birefringent and
  have polarization-dependent absorption. Here we present and
  demonstrate a simple scheme that allows compensating for these polarization
  effects. The scheme is demonstrated using a solid-state quantum
  memory implemented with an ensemble of rare-earth ions doped into a
  biaxial yttrium orthosilicate (Y$_2$SiO$_5$) crystal. Heralded
  single photons generated from a filtered spontaneous parametric
  downconversion source are stored, and quantum state tomography of
  the retrieved polarization state reveals an average fidelity of
  $97.5 \pm 0.4\%$, which is significantly higher than what is
  achievable with a \emph{measure-and-prepare} strategy.
\end{abstract}


\maketitle

Optical quantum memories~\cite{Lvovsky2009} allow storage of quantum
coherence and entanglement through the reversible mapping of quantum
states of light onto matter. They are an essential component of
long-distance quantum communication schemes based on quantum
repeaters~\cite{Sangouard2011} and of distributed quantum
networks~\cite{Kimble2008}. An important property of quantum memories
is the ability to operate with true single photons (i.e.~single-photon
Fock states). Single-photon qubits can be encoded using various
degrees of freedoms, such as temporal modes or
polarization. Polarization qubits are often employed, owing to the
simplicity in performing arbitrary single-qubit gates and projection
measurements. Yet, the storage of a qubit in an arbitrary state of
polarization is generally complicated by the fact that many quantum
memories are birefringent and have a polarization-dependent
absorption, e.g.~with solid-state devices. This not only leads to an
efficiency that depends on the state to be stored, but also introduces
a state dependent polarization transformation. Means to avoid this
drawback were investigated with memory-specific
approaches~\cite{Matsukevich2006a,Carreno2010,Specht2011a,Lettner2011,Viscor2011,Riedl2011a}
based on polarization-selection rules of suitably chosen atomic
transitions. Another approach, that can in principle work for any type
of quantum memory, is based on the embedding of two quantum memories
into the arms of an
interferometer~\cite{Cho2010,Zhang2011,England2011}, which inevitably
requires long-term interferometric stability.

In this Letter, we present a simple and stable storage
scheme for polarization qubits. We experimentally demonstrate our
scheme using birefringent crystals doped with rare-earth
ions~\cite{Tittel2010}. In recent years, this type of quantum memory
experienced rapid progress and key features were demonstrated, such as
long storage times~\cite{Longdell2005}, high-efficiency
storage~\cite{Hedges2010} and large temporal multi-mode
capacity~\cite{Usmani2010,Bonarota2011}. Operation at the single
photon level was demonstrated~\cite{Riedmatten2008} using the atomic
frequency comb protocol~\cite{Afzelius2009a}. This recently lead to
the storage of time-bin entangled photons generated from spontaneous parametric
downconversion~\cite{Clausen2011,Saglamyurek2011} and to the heralded
creation of entanglement shared between two
crystals~\cite{Usmani2011}. Here, we experimentally show that they can
also faithfully store polarization qubits encoded in heralded single
photons, despite their inherent birefringence and absorption
anisotropy. Our scheme can in principle work for all photon-echo based quantum
memories, independent of the material used for the implementation.

\begin{figure}[!h]
  \centering
  \begin{sansmath}
    \sffamily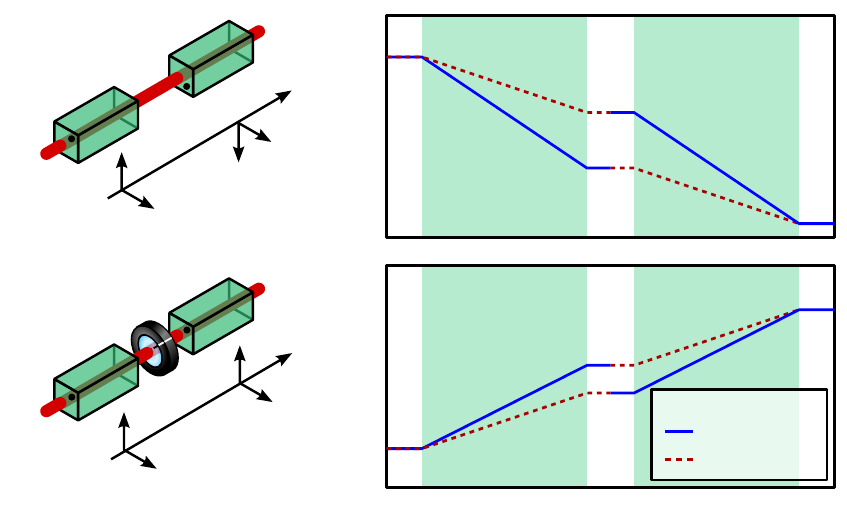
  \end{sansmath}
  \caption{Scheme for compensating birefringence and absorption
    anisotropy using two identical quantum memories
    $M_A$ and $M_B$. (a)~Light propagates along $+z$. The principal
    axes of the index of refraction $D_1$ and $D_2$ of each crystal
    are orthogonal to $z$, and $M_B$ is rotated $90^\circ$ around $z$
    with respect to $M_A$.  This arrangement creates a quantum memory
    that is polarization-preserving and has a constant efficiency for
    all input polarization states. (b)~Alternatively, a wave plate can
    be inserted between $M_A$ and $M_B$. (c)~Intensity transmitted
    along the two-memory arrangement of (a) and (b) for light
    polarized along $D_1$ or $D_2$. The total transmission is the same
    for both components, and hence for any linear combination.  (d)
    The accumulated optical phase for both polarization components
    along the two-memory arrangement.}
  \label{fig:method}
\end{figure}

Our scheme is illustrated in Fig.~\ref{fig:method}a using two
identical quantum memories $M_A$ and $M_B$ of length~$L$ each. Both
memories are placed along the $z$-axis such that their input facet is
oriented to contain two of the principal axes of the index of
refraction, which we label $D_1$ and $D_2$, and $M_B$ is rotated at
90$^{\circ}$ (around~$z$) with respect to $M_A$. Light propagates
along $+z$ and hits the memories at normal incidence. We assume that
the principal axes of absorption coincide with $D_1$ and $D_2$. We
denote $\alpha_1$ (or $\alpha_2$) the absorption coefficient along
$D_1$ (or $D_2$), and $d_1 = \alpha_1 L$ (or $d_2 = \alpha_2 L$) the
associated optical depth. The coincidence between the principal axes
of index of refraction and absorption is satisfied by many types of
quantum memories, such as rare earth-ions-doped crystals of high
symmetry~\cite{BornWolf}. Hence, the two linear polarizations states
parallel to $D_1$ and $D_2$ are eigenstates of each memory and form an
orthonormal basis that we use to decompose an arbitrary polarization
state.  Inside $M_A$, the components will experience different
absorption and phase shifts such that an arbitrary polarization input
state undergoes a non-unitary transformation stemming from the
combined effect of birefringence and absorption
anisotropy~\cite{Huttner2000}. Nevertheless, the two-memory
configuration is such that after passing through both $M_A$ and $M_B$,
both components experience the same total absorption and global phase
shift, and the arrangement is uniformly absorbing and
polarization-preserving. Using a rigorous decomposition of the whole
memory into infinitesimal longitudinal elements, we can show that the
memory efficiency of the forward re-emitted light can be the same for
all polarizations, and that the polarization is preserved (see
Appendix). The same conclusion holds with the configuration of
Fig.~\ref{fig:method}b where, instead of rotating $M_B$, a half-wave
plate inserted between $M_A$ and $M_B$ swaps the eigenstates (here,
the output polarization is preserved up to the swap operation).

We note that the high fidelity of our scheme is conditioned on the
interferometric stability of the distance between the two memories,
but only on the time scale of the storage time. This condition is
experimentally simple to satisfy, especially if the two memories are
mounted in a spatial configuration yielding common mode rejection of
vibration. In particular, this is easier to satisfy than the long-term
stability over the ensemble of all measurements that is required for the
scheme based on memories embedded in a Mach-Zehnder
interferometer~\cite{Cho2010,Zhang2011,England2011}.

\begin{figure}[!t]
  \centering
  \includegraphics[width=\linewidth]{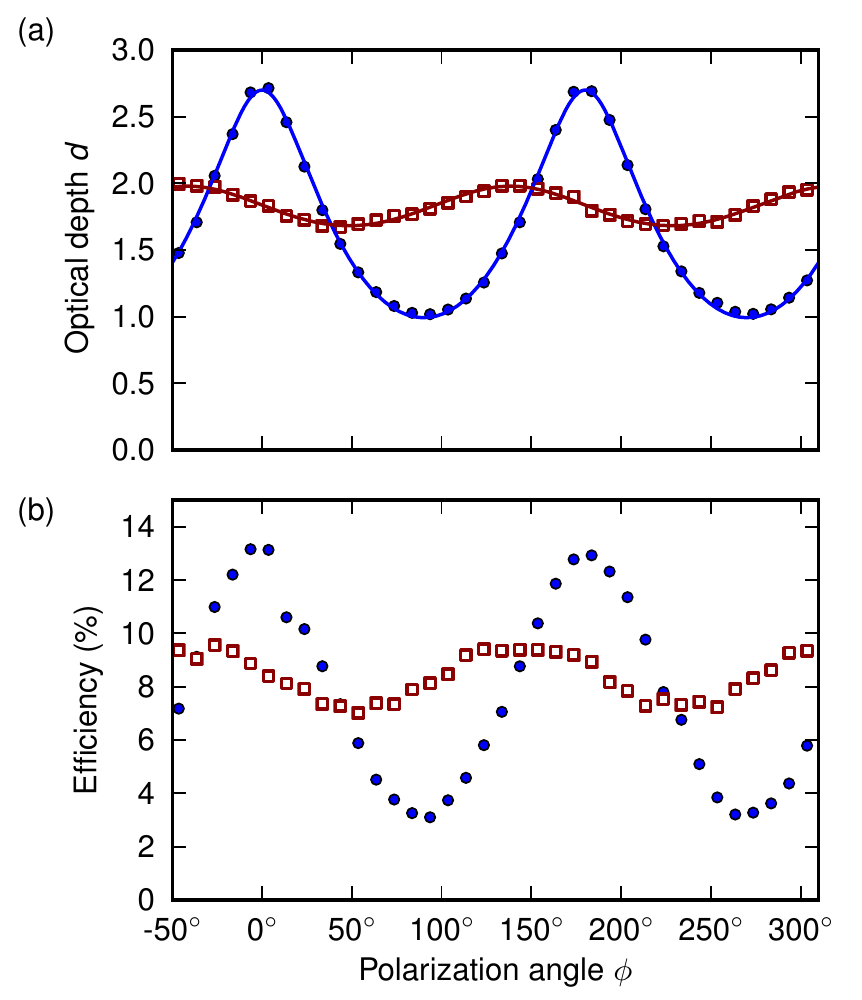}
  \caption{Measurements of (a)~optical depth and (b)~memory efficency
    of two \NdYSO{} crystals without~(\protect \bluecircle) and
    with~(\protect \redsquare) compensation scheme.  Incident light is
    linearly polarized, and the $x$-axis indicates the angle of the
    polarization with respect to the $D_1$ axis of the
    crystals. (a)~Without compensation the optical depth varies
    between 2.70(1) and 0.99(1), corresponding to propagation along
    the two optical extinction axes. With compensation the
    peak-to-peak variation is reduced to 16\% of the mean optical
    depth. Lines are fits to Eq.~\eqref{eq:odepth}. (b)~Efficiency of
    50~ns storage measured using laser pulses. With compensation the
    efficiency is almost independent of polarization.}
  \label{fig:depth_eff}
\end{figure}
We shall now present a series of measurements made with a pair of 1~cm long 
\NdYSO{} crystals for which we explicitly verified that the principal
axes of the index of refraction and of absorption coincide. We stress
that the host biaxial crystal is highly birefringent, with a
polarization beat length on the order of 100~\textmu m at 883~nm~\cite{Beach1990}. First, we show that the dependence of the optical depth on the
polarization of the light can be strongly attenuated. As a
consequence, the efficiency of the quantum memory is practically
constant. Second, we use the quantum memory for storage and retrieval
of heralded single photons with several polarizations, and obtain the
fidelity of the storage process using quantum state tomography.

We first measured the total optical depth $d$ on the $^4I_{9/2} \rightarrow{}
 ^{4}F_{3/2}$ transition at 883~nm of the pair of \NdYSO{} crystals,
cooled down to 3~K, while varying the polarization of the incident light. The
optical depth is obtained through $P_\text{out} =
P_\text{in}e^{-d}$, where $P_\text{in}$ (or $P_\text{out}$) is
the optical power before (or after) the two crystals. The two crystals
were initially in the configuration of Fig.~\ref{fig:method}b, but without the
wave plate between the crystals. The measured optical depth is shown
in Fig.~\ref{fig:depth_eff}a. Without the compensation
scheme, the absorption varies strongly with the linear polarization
state, as expected. The overall absorption for any linear polarization can be
calculated by decomposing the light into its components along the two
principal axes, such that the overall optical depth is given
by~\cite{Afzelius20101566},
\begin{equation}
  \label{eq:odepth}
  d = -\ln \left(e^{-d_1}\cos^2\phi + e^{-d_2}\sin^2\phi \right),
\end{equation}
where $d_1$ (or $d_2$) is the optical depth for polarization along
$D_1$ (or $D_2$), and $\phi$ the angle of the linear polarization with
respect to $D_1$. A fit of the data to Eq.~(\ref{eq:odepth}) results
in values $d_1 = 2.70(1)$ and $d_2 = 0.99(1)$. These
values are consistent with previous measurements on a
single crystal~\cite{Usmani2010}. We then repeated the measurement
using the compensation scheme of Fig.~\ref{fig:method}b. 
With compensation, the minimum and maximum values of the optical
depth, obtained from another fit to Eq.~(\ref{eq:odepth}), were found
to be 1.68(1) and 1.98(1). We believe that the residual variation can
be attributed to birefringence induced by the windows on the cryostat,
slight misalignment of the crystals with respect to each other, or 
unequal crystal lengths. Additionally, it is possible that the
retardance and orientation of the half-wave plate deviated from the
ideal values.

The optical depth is important for the efficiency of the quantum
memory. In the case of the atomic frequency comb, the efficiency
varies as $\eta \propto \tilde{d}^2 e^{-\tilde{d}}$, where
$\tilde{d}=d/F$ is the average optical depth of the comb, including a
dependence on the comb finesse $F$~\cite{Afzelius2009a}. This means
that variations in the optical depth translate directly to variations
in the memory efficiency. We verified this using the same setup as for
the measurement of the optical depth, but now the crystals were
prepared as 120~MHz wide atomic frequency combs with a preprogrammed
storage time of 50~ns (see~\cite{Clausen2011,Usmani2011} for
details). Using laser pulses of roughly Gaussian shape with 18~ns full
width at half maximum, we measured the memory efficiency, that is, the
ratio of the area of the retrieved pulse to that of the input pulse,
as a function of polarization. Figure~\ref{fig:depth_eff}(b) shows
that the efficiency follows the same trend as the optical depth,
varying from 3\% to 13\% without, and from 7\% to 9.5\% with
compensation.

\begin{figure}[!t]
  \centering
  \begin{sansmath}
    \sffamily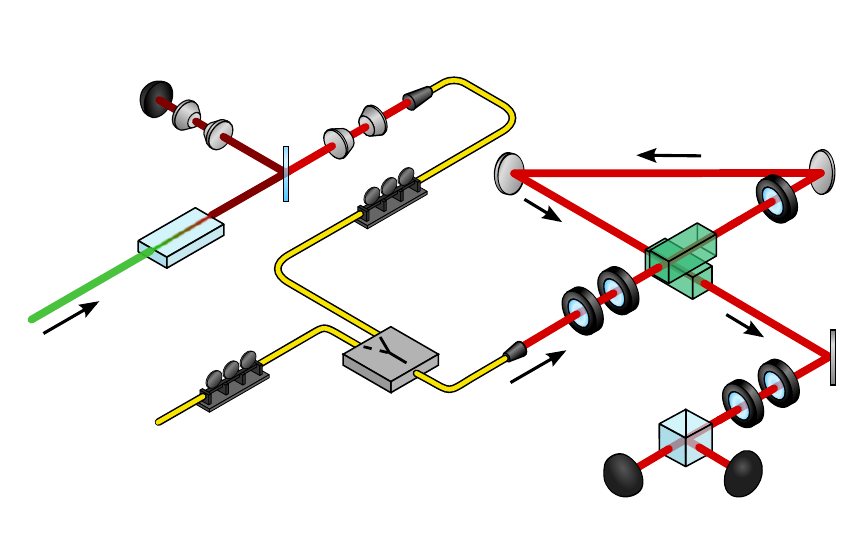
  \end{sansmath}
  \caption{Experimental setup for polarization qubit tomography. Pairs
    of photons are generated by spontaneous parametric down-conversion
    in a periodically poled KTP waveguide pumped with 4~mW of light
    from a continous-wave laser at 532~nm, separated by a dichroic
    mirror (DM) and strongly filtered. The detection of a photon at
    1338~nm on a superconducting nanowire single photon detector
    heralds the presence of a photon at 883~nm. The latter is in
    resonance with the quantum memories $M_A$ and $M_B$ and acts as a
    polarization qubit. The initial state of the qubit is prepared
    using a fiber-based polarization controller (PC) and two
    waveplates ($\lambda/2$ and $\lambda/4$). An effective half-wave
    plate, made of two quarter-wave plates between the crystals,
    compensates for their anisotropy. Finally, the polarization state
    of the retrieved photon is analyzed using two more waveplates
    ($\lambda/4$ and $\lambda/2$) and a polarizing beamsplitter (PBS)
    with a silicon avalanche photo diode at each output. Another
    polarization controller is used to ensure that the quantum memory
    is always prepared with the same polarization, independent of the
    polarization state of the photons to be stored.}
  \label{fig:setup}
\end{figure}

Let us now turn to the storage and retrieval of polarization
qubits. The complete setup for this purpose is shown in
Fig.~\ref{fig:setup}. The experimental cycle consists of two phases of
15~ms each. In the first phase the inhomogeneous absorption of the two
crystals is shaped into an atomic frequency comb by optical pumping
during 11~ms (see~\cite{Clausen2011} for details). A waiting time of
4~ms is added to avoid fluorescence from atoms left in the excited
state. In the second phase the storage, retrieval and analysis of
polarization qubits, encoded on heralded single photons, is
performed. Our photon source is described in detail
elsewhere~\cite{Clausen2011,Usmani2011}. Photon pairs are generated by
spontaneous parametric down-conversion in a periodically poled
waveguide. The signal and idler photons with wavelengths 883~nm and
1338~nm, respectively, are separated and strongly filtered to match
the bandwidth of the quantum memories. The detection of an idler
photon heralds the presence of a signal photon, which will act as the
polarization qubit. The signal photon then passes through a
fiber-based polarization controller, a half-wave plate and a
quarter-wave plate to prepare the state of the qubit to be stored. The
output of the quantum memory is sent towards a polarization analyzer
consisting of a quarter-wave plate, a half-wave plate, a polarizing
beamsplitter and two silicon-based single photon detectors. The light
that prepares the atomic frequency comb passes through the same
waveplates as the single photons. We adjust the polarization of the
preparation light using a fiber-based polarization controller, such
that it always has the same polarization, independent of the
polarization of the photons to be stored.

\begin{table}
  \centering
  \begin{ruledtabular}
    \begin{tabular}{ccc}
      Input State & Fidelity & $\bar{g}^{(2)}_{si}$ \\
      \hline
      $\ket{H}$ & 99.3(6)\% & 7.6(3) \\
      $\ket{V}$ & 97(1)\% & 6.0(3) \\
      $\ket{L}$ & 97.7(6)\% & 9.4(3) \\
      $\ket{+}$ & 95(1)\% & 8.0(3) \\
      $\alpha\ket{H} + \beta\ket{V}$ & 98.7(9)\% & 9.2(3) \\
    \end{tabular}
  \end{ruledtabular}
  \caption{Results of the measurements performed on the stored and
    released single photons for various input states. The fidelity was
    found via tomographic state-reconstruction using a maximum
    likelihood method, and the errors estimated via Monte-Carlo
    simulation. Additionally, the cross-correlation between signal and
    idler photons $\bar{g}^{(2)}_{si}$, averaged over measurement settings
    and detectors, indicates the quantum character of the process. The
    fifth input state is obtained by sending $\ket{H}$ through a
    quarter-wave plate oriented such that $\alpha=(1+i\sqrt{2})/2$
    and $\beta=1/2$.}
  \label{tbl:results}
\end{table}

To show that our compensation method allows for faithful storage of
polarization qubits, we performed quantum state
tomography~\cite{Altepeter2006} on a set of five different input
states. The results are shown in Table~\ref{tbl:results}.  
For each tomographic reconstruction we performed
measurements along the three principal axes of the Poincar\'e sphere,
and all detections were conditioned on the simultaneous detection
of an idler photon.

Based on the measured number of coincidences, we reconstructed the
density matrix of the retrieved qubit by maximum likelihood
estimation~\cite{Altepeter2006}. The fidelity with respect to the
input state is shown in Table~\ref{tbl:results}. We also estimated the
uncertainty on the fidelity using a Monte-Carlo simulation based on
the Poissonian statistics of the detection events. In all cases we
find fidelities $F \gtrsim 95\%$, which significantly surpasses the
limit of $2/3$ of any classical memory based on a \emph{measure-and-prepare}
strategy~\cite{Massar1995}. This proves that we have implemented a
quantum memory that preserves arbitrary states of polarization to a
high degree. We attribute the deviation from unit fidelity to
imperfections in the state preparation, anisotropy compensation and
analysis, caused by non-ideal wave plates.

Finally, we investigated the quantum nature of the storage
process. First, a measurement of the zero-time auto-correlation
function of the heralded signal photon (before storage) yielded
$g^{(2)}_{s|i} < 0.06$, which confirms the single-photon nature of the
polarization qubit to be stored. The non-classicality of the photon
retrieved from the memory can then be revealed with another
measurement, namely the zero-time intensity cross-correlation
$g^{(2)}_{si}$ between signal and idler fields. Specifically, assuming
auto-correlations $g^{(2)}_x$ for signal and idler fields of $1 \leq
g^{(2)}_x \leq 2$, where $x=$`$s$' for signal and `$i$' for idler, the
signature of non-classicality between both fields becomes
$g^{(2)}_{si} > 2$~\cite{Kuzmich2003}.  We measured values between
$\bar{g}^{(2)}_{si} = 6.0(3)$ and $\bar{g}^{(2)}_{si} = 9.4(3)$ for
all the stored polarization states, confirming the quantum nature of
the storage and retrieval process. The lowest cross-correlation value
allows to upper bound the auto-correlation of the retrieved signal
photon to $g^{(2)}_{s|i} \leq 0.61(3)$, which is still below the
classical threshold of~1 (the upper bound is obtained by assuming that
the source is exactly described by a two-mode squeezed
state~\cite{Usmani2011}).  We note that the relatively large increase
of $g^{(2)}_{s|i}$ before and after storage is almost entirely due to
an experimental artifact associated with the continuous wave operation
of our source of photon pairs~\cite{Usmani2011}, and not to a
detrimental effect stemming from the memory itself.

To conclude, we have experimentally demonstrated a scheme that allows
the faithful storage of polarization qubits encoded into true single
photons using a material that is birefringent and has anisotropic
absorption.  We note that the efficiency of a photon-echo based
quantum memory with re-emission in the forward mode is limited to
54\%~\cite{Sangouard2007,Afzelius2009a}. To overcome this limitation,
and possibly reach 100\% efficiency, one possibility is to use the
impedance-matched quantum memory scheme in which a forward-emitting
quantum memory is placed between two-mirrors with reflectivity chosen
such that all the incident light can be absorbed and
re-emitted~\cite{Afzelius2010a}. Our scheme thus has the potential for
demonstrating a high-efficiency solid-state quantum memory that is
compatible with polarization qubits.  It is particularly well-suited
for rare-earth-ions-doped crystals and greatly extends their range of
application. For example, the storage of both polarization and
temporal modes leads to new interesting possibilities, such as the
quantum storage of hyperentangled photons~\cite{Barreiro2005}.

\begin{acknowledgements}
  We thank Imam Usmani, Nicolas Sangouard and Philippe Goldner for stimulating
  discussions.  We acknowledge support by the Swiss NCCR Quantum
  Photonics, the Science and Technology Cooperation Program
  Switzerland-Russia, as well as by the European projects QuRep and
  ERC-Qore. F.B. was supported in part by le Fond Qu\'eb\'ecois de la
  Recherche sur la Nature et les Technologies.
\end{acknowledgements}

We note that related results have been obtained in other
groups~\cite{Saglamyurek2011a,Gundogan2012,Zhou2012}.

\appendix*
\section{Appendix}

We here present a detailed theoretical description of our scheme for compensating birefringence and anisotropic absorption in quantum memories. We represent the polarization input state in terms of a Jones vector $\ket{\Psi_i}$ that is transformed to an output state $\ket{\Psi_f}=\mathbf{M}\ket{\Psi_i}$, where $\mathbf{M}$ is a 2$\times$2 square complex matrix.

We use the theory derived in \cite{Huttner2000}, which treats the effects of birefringence (or  polarization-mode dispersion, PMD) and polarization-dependent loss (PDL). In \cite{Huttner2000} it is shown that any transformation $\mathbf{M}$ of the input state induced by propagation through a memory having PMD and PDL effects, can be decomposed into a product of an unitary matrix $\mathbf{U}$ describing an effective PMD and an Hermitian positive matrix $\mathbf{T}$ describing an effective PDL: 
\begin{equation*}
\mathbf{M}=\mathbf{TU}.
\end{equation*}

In the most general case when the principal axes of the PDL and PMD do not coincide, $\mathbf{T}$ and $\mathbf{U}$ do not commute \cite{Huttner2000}. For rare-earth doped crystals, in most materials commonly used, the principal axes of the absorption coefficient and the index of refraction do coincide due to symmetry considerations. This is the case, for instance, when the dopant site has an axial symmetry. In our crystal, Nd:Y$_2$SiO$_5$, the neodymium ions are in a crystallographic site of low symmetry $C_1$ and Y$_2$SiO$_5$ is a biaxial crystal. The principal axes of absorption and birefringence could thus be oriented differently. The data presented in the Letter shows, however, that the principal axes do coincide (at least in the plane perpendicular to the crystallographic b-axis). We will thus consider that $\mathbf{U}$ and  $\mathbf{T}$ are diagonal in the same basis. The principal axes of birefringence in Y$_2$SiO$_5$ are denoted $D_1$ and $D_2$ in the Letter. 

The matrices $\mathbf{T}$ and $\mathbf{U}$ can then be written as
\begin{equation*}
\mathbf{U}=
 \begin{pmatrix}
  1 & 0\\
  0 & e^{i\phi}
 \end{pmatrix}\quad \text{and}\quad
\mathbf{T}=
 \begin{pmatrix}
  e^{-d_1/2} & 0\\
  0 & e^{-d_2/2}
 \end{pmatrix},
\end{equation*}

\noindent where $d_1$ and $d_2$ are the optical depths along the
principal axes $D_1$ and $D_2$ and $\phi = k \Delta n L$ is the phase
accumulated along axis $D_2$, relative to $D_1$, due to birefringence
$\Delta n = n_2 - n_1$ (that is the difference between the indices of
refraction along $D_2$ and $D_1$). It is now easily shown that putting
two crystals in series, where the second crystal is rotated 90 degrees
with respect to the first, results in the identity transformation
multiplied by a global polarization-independent absorption (loss)
\begin{equation*}
\mathbf{R}(-90^{\circ})\mathbf{TU}\mathbf{R}(90^{\circ})\mathbf{TU}=e^{-(d_1+d_2)/2}
e^{i\phi}\,\mathbf{I}.
\end{equation*}

\noindent Here $\mathbf{R}(\alpha)$ is the usual 2$\times$2 rotation matrix, $\alpha$ is the angle of rotation, and $\mathbf{I}$ is the identity matrix. Note that in terms of intensity the fraction of absorbed light is $1-\exp[-(d_1+d_2)]$.

We have so far shown that the propagation through the two absorbing crystals preserves the polarization state, and that light is uniformly absorbed (or mapped onto) the crystal. For quantum memories it is of course also important to show that the light state can be retrieved efficiently and that the polarization state of the retrieved light is preserved as well.

We model the quantum memory as a ``sum over all trajectories"~\cite{Sangouard2007,Afzelius2009a,Afzelius2010a}. Each trajectory corresponds to a conversion from the input mode to the output mode at a particular position $z$ in the absorbing medium. Let us assume that the polarization of the input mode corresponds to a polarization eigenvector of the memory, such that its polarization does not rotate through propagation. Moreover, we consider the case where the input and output modes propagate in the same ``forward" direction. In this situation, the square root of the memory efficiency, $\sqrt{\eta}$, can be calculated as (see Eq. (A19) in \cite{Afzelius2009a})
\begin{equation}
\sqrt{\eta} = \int \limits_0^L e^{-\alpha z/2} \alpha dz
e^{-\alpha (L-z)/2}=\alpha L
e^{-\alpha L/2} = d e^{-d/2},
\label{singlepass_eta}
\end{equation}
where $\alpha$ is the absorption coefficient, $L$ the memory length  and $d=\alpha L$ is the optical depth. 
The equation above can be understood in the following way. For each trajectory there is the probability amplitude for the input photon to reach point $z$ in the medium ($e^{-\alpha z/2}$), the probability amplitude for the photon to be absorbed in $z$ and re-emitted later ($\alpha dz$) and the probability amplitude for the output photon to reach the end of the medium ($e^{-\alpha (L-z)/2}$). All trajectory amplitudes can be added coherently in the far field, leading to the final efficiency formula for ``forward" memory read-out. For the specific case of the atomic frequency comb protocol, the reasoning is the same, but the optical depth is replaced with $\tilde{d} = d/F$, where $F$ is the finesse of the comb~\cite{Afzelius2009a}.

The well-known result of Eq.~\ref{singlepass_eta} applies to a single polarization mode. The calculation in our case is more general but nevertheless straightforward. The extension to two polarization modes for one crystal can be done as follows:
\begin{equation*}
\mathbf{M}'=\int \limits_0^L
\mathbf{T}(L-z)\mathbf{U}(L-z)
 \begin{pmatrix}
  \alpha_1 dz & 0\\
  0 & \alpha_2 dz
 \end{pmatrix}
 \mathbf{T}(z)\mathbf{U}(z).
\end{equation*}
\noindent where the diagonal matrix in the middle accounts for the conversion efficiency for the two polarization modes. This equation can be simplified extensively:
\begin{widetext}
\begin{eqnarray*} 
\mathbf{M}'=\int \limits_0^L
\begin{pmatrix}
  e^{-(L-z)\alpha_1/2} & 0\\
  0 & e^{i\phi}e^{-(L-z)\alpha_2/2}
 \end{pmatrix}
  \begin{pmatrix}
  \alpha_1 dz & 0\\
  0 & \alpha_2 dz
 \end{pmatrix}
 \begin{pmatrix}
  e^{-z\alpha_1/2} & 0\\
  0 & e^{i\phi} e^{-z\alpha_2/2}
 \end{pmatrix} = \\ \int \limits_0^L
 \begin{pmatrix}
  e^{-(L-z)\alpha_1/2} \alpha_1 dz e^{-z\alpha_1/2} & 0\\
  0 & e^{i\phi} e^{-(L-z)\alpha_2/2} \alpha_2 dz e^{-z\alpha_2/2}
 \end{pmatrix} = 
  \begin{pmatrix}
  d_1 e^{-d_1/2} & 0\\
  0 & e^{i\phi} d_2 e^{-d_2/2}
 \end{pmatrix}.
\end{eqnarray*}
\end{widetext}

\noindent The effect of both crystals can now be evaluated by summing the amplitudes of the following two possibilities; a photon is stored in the first crystal and propagates through the second crystal, or the vice versa
\begin{multline*}
\mathbf{R}(-90^{\circ})\mathbf{M}'\mathbf{R}(90^{\circ}) \mathbf{M}
+\mathbf{R}(-90^{\circ})\mathbf{M}\mathbf{R}(90^{\circ}) \mathbf{M}' \\
= e^{i\phi} (d_1+d_2) e^{-(d_1+d_2)/2}\, \mathbf{I}.
\end{multline*}
\noindent We conclude that the total transformation of the input to output mode preserves the polarization state and the efficiency is given by the usual formula with an effective optical depth $d_1+d_2$, as one could have expected. The efficiency in this configuration would thus ultimately be limited to 54\% for $d_1+d_2 \approx 2$ \cite{Afzelius2009a}. This assumes, of course, that there is no decoherence during the time spent in the memory. Decoherence can, however, be included by simply multiplying the equation above with the appropriate decoherence factor~\cite{Sangouard2007}.

It has been shown previously that efficiencies approaching 100\% can be achieved in the so-called ``backward" read out configuration \cite{Afzelius2009a}. Using our scheme, however, it is clear that one cannot in general hope to achieve unit efficiency in the backward direction. This is due to the fact that different trajectories with penetration depth $z$ accumulate different phases, due to the birefringence. In the special case of negligible birefringence ($\phi \approx 0$), but with anisotropic absorption, one can still achieve unit efficiency in the backward read out configuration. The calculation leading to this result is similar to the one presented above.

Another way of reaching 100\% efficiency is to put the memory in a moderate-finesse cavity, based on the concept of impedance-matching \cite{Afzelius2010a}. That scheme is perfectly suitable for our anisotropy compensation scheme, since for each cavity round trip the memory works in forward emission mode. The results presented above thus apply to this type of memory. There exist thus a method to extend the ideas presented here to unit efficiency quantum memories.


\end{document}